\pdfoutput=1

\documentclass{article}

\usepackage{spconf,amsmath,graphics,graphicx}
\usepackage{amsfonts}
\usepackage{subcaption}
\usepackage{xcolor}
\usepackage{mwe} 

\newcommand{\cmmnt}[1]{}

\definecolor{airforceblue}{rgb}{0.36, 0.54, 0.66}

\title{Deep networks to automatically detect\\ late-activating regions of the heart}
\name{
\parbox{\linewidth}{
\centering
Jiarui Xing$^{ a}$ \qquad
Sona Ghadimi$^{ c}$ \qquad 
Mohammad Abdishektaei$^{ c}$ \qquad \\
Kenneth C. Bilchick$^{ d}$ \qquad
Frederick H. Epstein$^{ c}$ \qquad
Miaomiao Zhang$^{ a, b}$
}}

\address{
\parbox{\linewidth}{
\centering
$^{a}$ Department of Electrical and Computer Engineering, University of Virginia, USA \\
$^{b}$ Department of Computer Science, University of Virginia, USA \\
$^{c}$ Department of Biomedical Engineering, University of Virginia Health System, USA\\
$^{d}$ School of Medicine, University of Virginia Health System, USA
}}
\begin{document}
%
\maketitle



%
\begin{abstract}
This paper presents a novel method to automatically identify late-activating regions of the left ventricle from cine Displacement Encoding with Stimulated Echo (DENSE) MR images. We develop a deep learning framework that identifies late mechanical activation in heart failure patients by detecting the Time to the Onset of circumferential Shortening (TOS). In particular, we build a cascade network performing end-to-end (i) segmentation of the left ventricle to analyze cardiac function, (ii) prediction of TOS based on spatiotemporal circumferential strains computed from displacement maps, and (iii) 3D visualization of delayed activation maps. Our approach results in dramatic savings of manual labors and computational time over traditional optimization-based algorithms. To evaluate the effectiveness of our method, we run tests on cardiac images and compare with recent related works. Experimental results show that the proposed approach provides fast prediction of TOS with improved accuracy. 
\end{abstract}
%

%
\section{Introduction}
Cardiac resynchronization therapy (CRT) is widely used to treat cardiac  conduction system disorders, such as left bundle branch block (LBBB) and  intrinsic  myocardial  diseases~\cite{abraham2002cardiac, lindenfeld_effects_2007, moss2009cardiac}. However, standard CRT suffers from a high non-response rate (approximately 40\%)~\cite{chung2008results, exner2012contemporary}. Implanting the CRT left ventricle lead at an area with delayed activation may significantly decrease the non-response rate~\cite{bilchick2014impact, ramachandran2015singular}. Therefore, it is critical to develop a method that accurately measures left ventricular activation time to improve CRT lead site selection and its response rate.

Cine DENSE MR imaging is an accurate and reproducible method for imaging regional myocardial displacement and strain~\cite{aletras1999dense, kim2004myocardial}. Studies have shown that late mechanical activations can be effectively measured on circumferential strains of tissue displacements imaged by DENSE~\cite{bilchick2014impact, ramachandran2015singular}. Other examples of strain-based cardiac activation estimation algorithms can be found in~\cite{kvaale2019detection, wyman1999mapping}. These methods measure the activation at each location of the left ventricle separately, which introduce artifacts of spatial inconsistencies. To alleviate this problem, a recent work~\cite{auger2017imaging} developed a semi-automatic algorithm based on active contour models with strain vectors concatenating information of all myocardial segments~\cite{auger2017imaging}. However,the estimation of~\cite{auger2017imaging} becomes unstable when the signal-to-noise ratio is low. Moreover, an additional requirement of fine-tuning the model parameters case by case greatly hinders its applicability in real clinical settings. 




To address the issue above, this paper introduces a deep learning-based method that automatically predicts delayed activation time of ventricular arrhythmias (a.k.a. TOS) without the need of manual inputs and parameter-tuning. More specifically, we develop a cascade deep learning framework that integrates multiple functionalities, including automatic segmentation of left ventricles from cardiac images, TOS prediction by analyzing spatiotemporal strains through convolutional neural networks (CNN) , and 3D visualization of activation maps. In contrast to traditional approaches, our method provides more reliable and stable estimation with significantly reduced computational time and intensive labor. To evaluate the effectiveness of our approach, we run tests on DENSE MR images and compare with the state-of-the-art strain-based activation time estimation algorithm~\cite{auger2017imaging}.

\section{Method}
In this section, we present a new framework that automatically detects and reconstructs 3D myocardial activation maps of left ventricles from cine DENSE MRIs. We develop a cascade network where an automatic segmentation of left ventricles is followed by abnormal activation detection of the ventricles in patients with cardiac arrhythmias. Given the input of DENSE images, our model learns a way to reconstruct 3D visualization maps of activation time in an end-to-end fashion. Before introducing our network architecture, we will first review the strain analysis of DENSE imaging~\cite{Ghadimi2020myocardial}. 

\subsection{Circumferential Strain Analysis}
\begin{figure}[ht!]
\centering
\begin{subfigure}{0.4\linewidth}
  \centering
  \includegraphics[width=\textwidth]{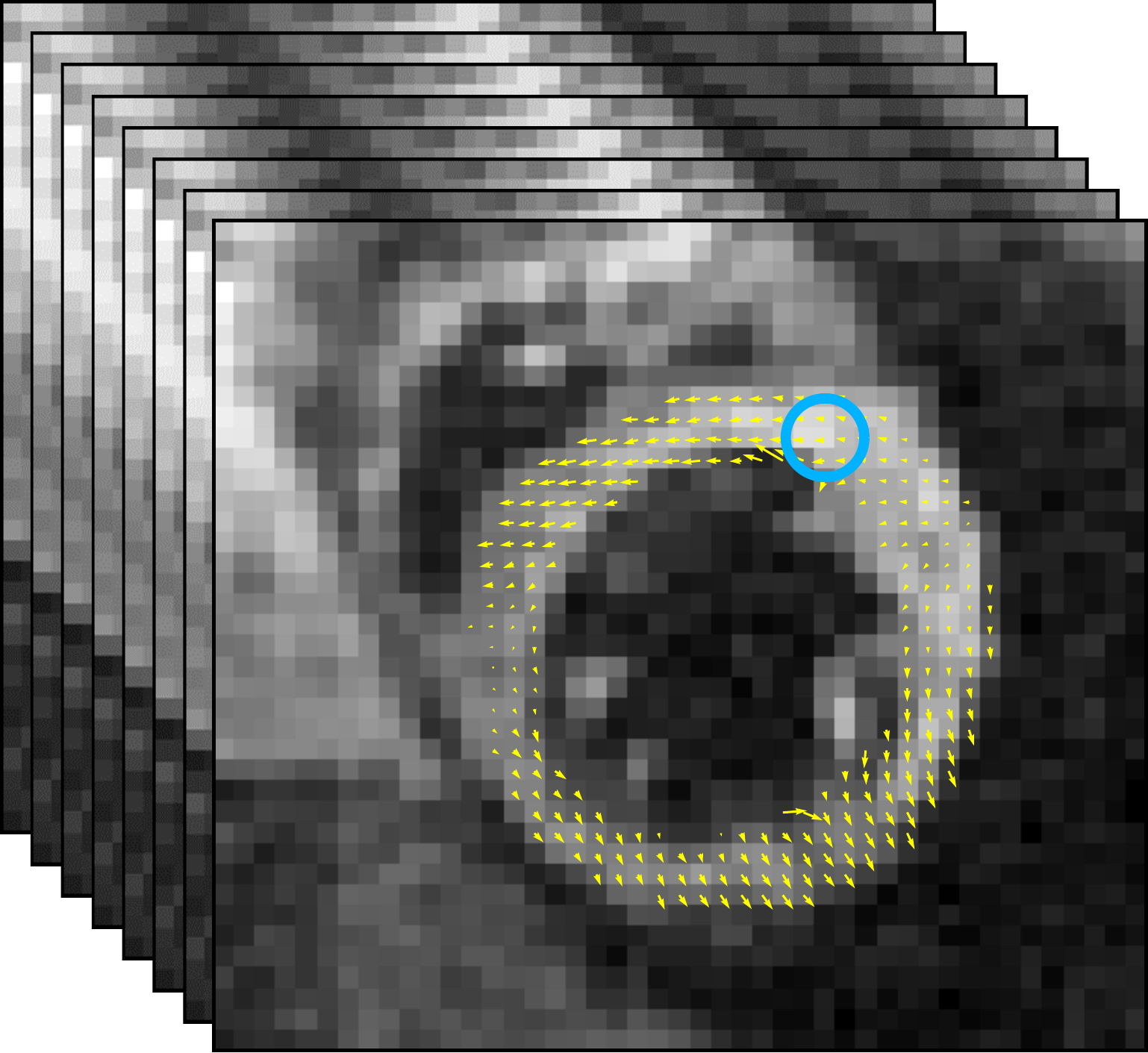}
    \caption{}
  \label{fig:disp}
\end{subfigure} %
\hspace{.7cm}
\begin{subfigure}{.4\linewidth}
  \centering
  \includegraphics[width=\textwidth]{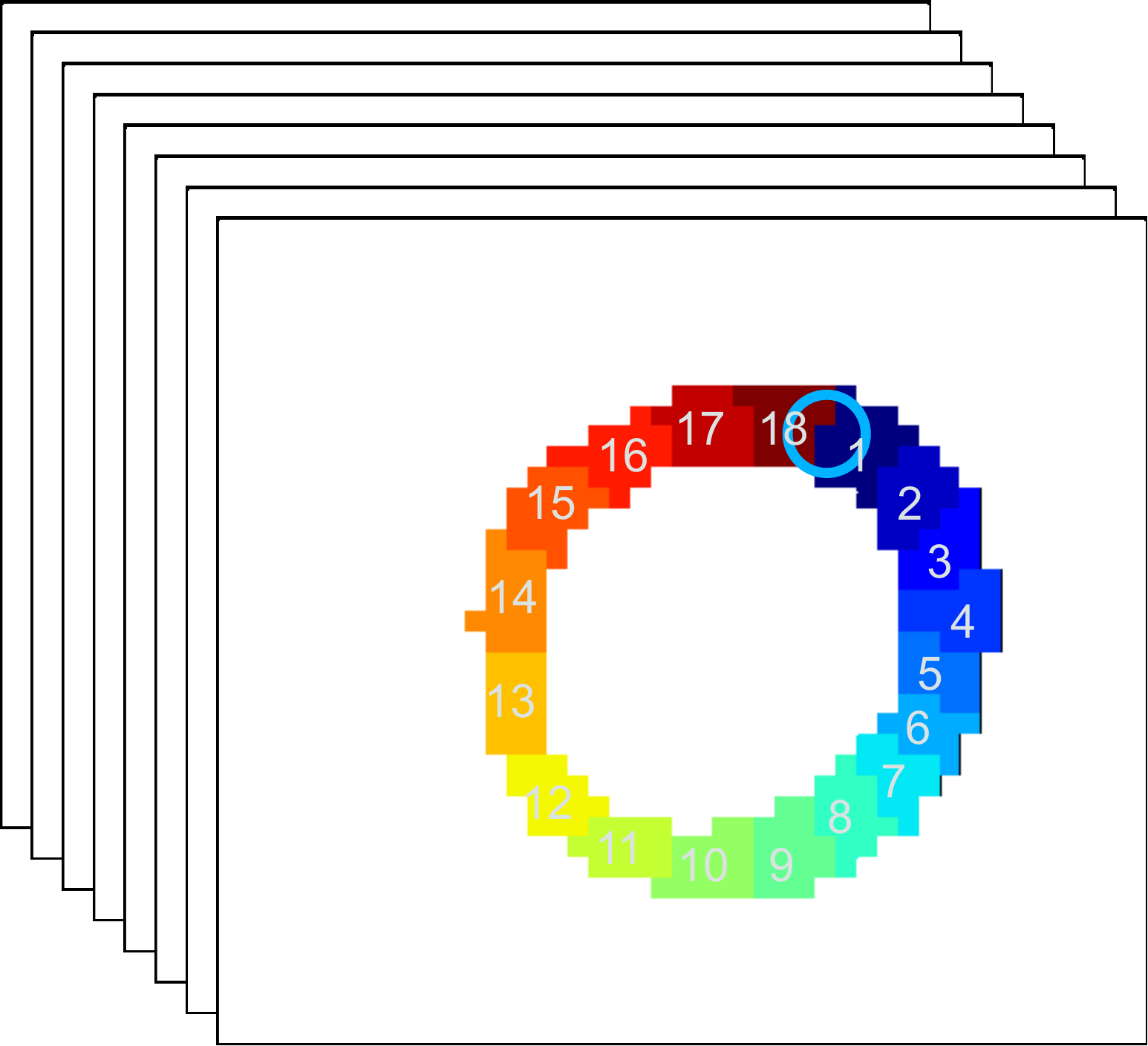}
  \caption{}
  \label{fig:segments}
\end{subfigure} \\
\begin{subfigure}{.5\linewidth}
  \centering
  \includegraphics[width=\textwidth]{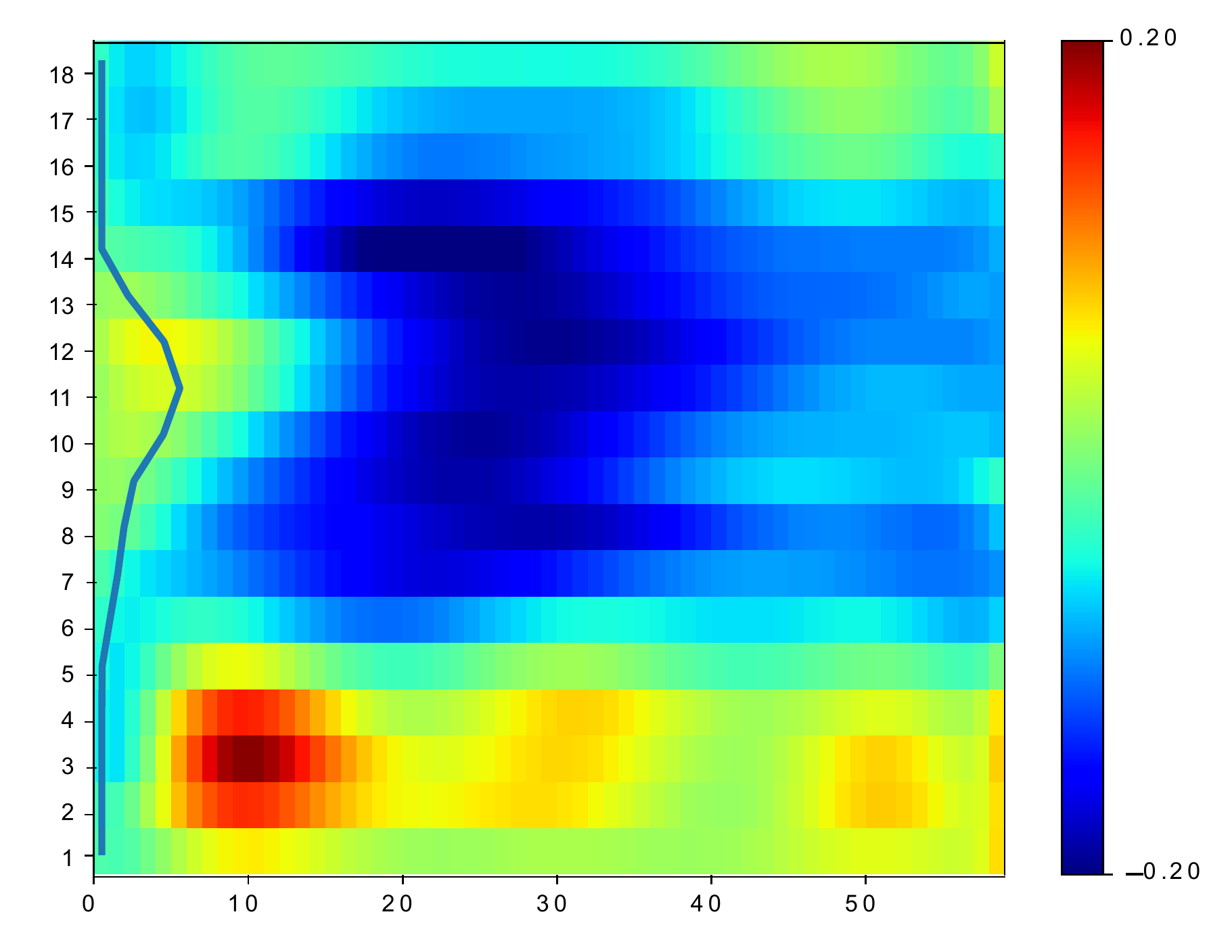}
    \caption{}
  \label{fig:strainMat_TOS}
\end{subfigure} %
\caption{Example of (a) temporal DENSE magnitude MRIs overlaid with displacement maps; (b) myocardial segments of left ventricles with strain values (blue color indicates contraction vs. red color indicates expansion); (c) strain matrix and its corresponding TOS curve. Blue circles in (a) and (b) indicate intersections of left and right ventricles.}
\label{fig:background}
\end{figure}

Circumferential strain has been shown to represent the myocardial contraction (with negative value) along the circular outline in the short axis~\cite{budge2012mr}. Advantages of analyzing strain values include (i) significantly reducing the computational complexity of our network input from displacements (vector-valued spaces $V$) to strain images (real-valued functions $S$), (ii) more robustness to body motion artifacts that occurred in original signals, and (iii) its effectiveness of dyssynchrony quantification~\cite{budge2012mr} as well as high reproducibility for validation~\cite{young2012generalized}. Following a recent work of~\cite{dandel2009strain}, we estimate the left ventricle activation time based on strains computed from the displacement maps. Examples of MR images overlaid with displacement maps over time are shown in Fig.~\ref{fig:disp}. 

Let $\Omega = \mathbb{R}^d / \mathbb{Z}^d$ be a $d$-dimensional torus domain with periodic boundary conditions. Given a $d$-dimensional displacement map $u(x):\Omega \rightarrow \mathbb{R}^d$, a Jacobian matrix $D^{d \times d}$ of $u$ at each spatial location is 

\begin{equation}
    D = \begin{pmatrix}
        \Delta u_1^1 & \cdots & \Delta u_1^d \\ 
        \vdots & \ddots & \vdots \\ 
        \Delta u_d^1 & \cdots & \Delta u_d^d
        \end{pmatrix}, 
    \nonumber
\end{equation}
where $\Delta u_i^j=\frac{\mathbf{d} u_i}{\mathbf{d} x_j}$, with $i, j \in \{1, \cdots, d\}$. A strain tensor $E$ is then computed as $E = \frac{1}{2}(D^T D-I)$, where $T$ denotes a matrix transpose and $I$ is a $d \times d$ identity matrix. This paper focuses on the circumferential strain, which is the component of strain along the myocardium. 

In practical clinical applications, the muscle area of the left ventricle is often divided into a variable number of segments~\cite{cerqueira2002american,american2002standardized}. In this paper, we divide each of the basal, mid-plane, and apical slices of short-axis view into $18$ segments (as shown in Fig.~\ref{fig:segments}). The starting location of the segments is at the upper intersection of the left and right ventricles (labeled by a blue circle in Fig.~\ref{fig:disp} and Fig.~\ref{fig:segments}). Each segment covers the same angle with respect to the center of left ventricles. A $18$-dimensional strain vector is determined from the full-resolution strain of left ventricle, with strain value extracted at the center of each segment for each slice. To facilitate the automatic estimation of TOS, we arrange the temporal strain data into a 2D matrix consisting of all strain vectors over a time-sequence of cardiac phases. Example of spatiotemporal strain matrix overlaid with circumferential TOS curve is shown in Fig.~\ref{fig:strainMat_TOS}. Note that we set the starting segment at the upper intersection of the left and right ventricles across all experiments.    

\subsection{Network Architecture}
\begin{figure*}[!ht]
    \centering
    \includegraphics[width=0.95\textwidth]{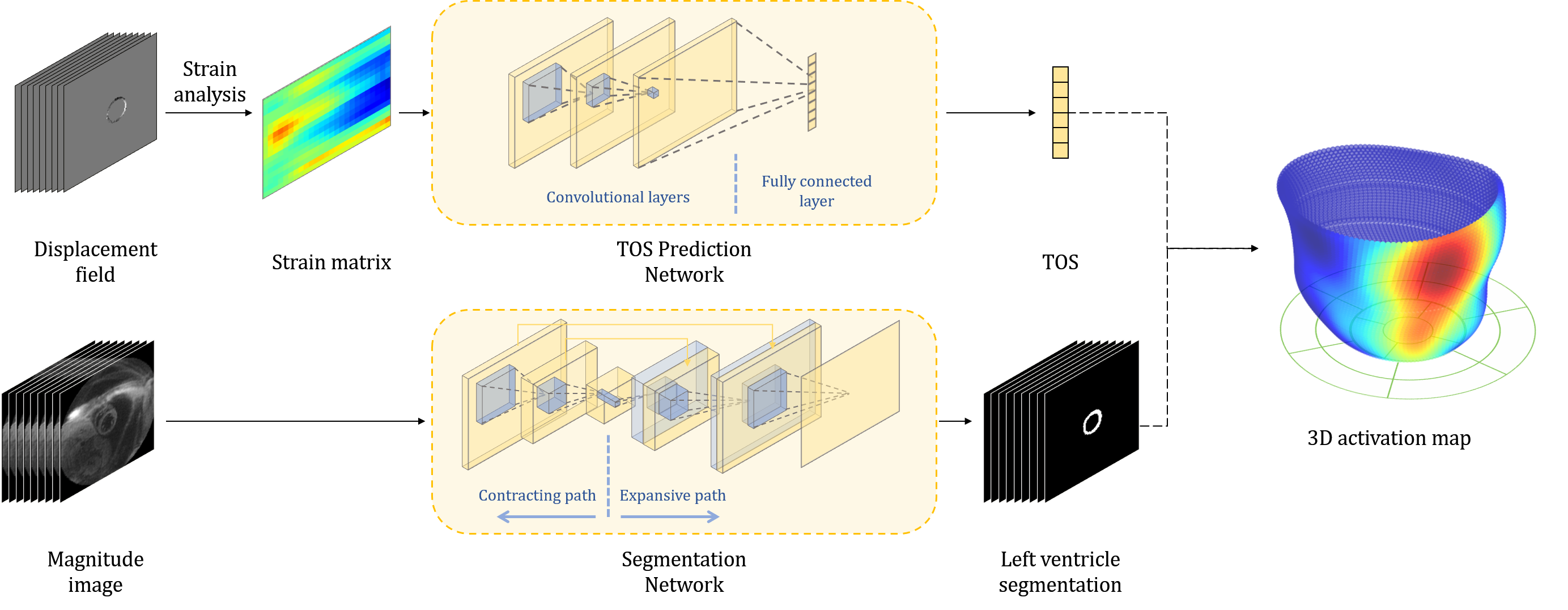}
    \caption{Illustration of our approach: a unified network of segmentation and delayed activation time prediction.}
    \label{fig:pipeline}
\end{figure*}

Now we are ready to introduce the network architecture (as illustrated in Fig.~\ref{fig:pipeline}. There are three major components in our model, including an automatic segmentation network providing segmentation maps for left ventricles from cine DENSE MRIs, a CNN-based network predicting TOS curves, and 3D reconstruction and visualization of delayed activation maps. Details are elaborated in the following sections. 

\noindent\textbf{Segmentation Network.} We employ a 2D segmentation neural network proposed in~\cite{Ghadimi2020myocardial} for left ventricle segmentation. The 2D U-Net networks utilize the structure presented by Ronneberger et al.~\cite{ronneberger2015u} with modifications for better performance. More specifically, we use dilated convolutional layers in the contracting path of standard U-Net, which increases the receptive field size without increasing the number of parameters and shows improved performance in our experiments. Our loss function is the summation of the weighted pixel-wise cross entropy and soft dice loss. During the training process, data augmentation on-the-fly is performed by applying random translations, rotations and scalings followed by B-spline-based transformations on input images and their corresponding ground-truth label maps at each iteration. To improve the accuracy and smoothness of the segmented contours, each testing image is rotated $9$ times with an interval of $40$ degrees. The corresponding output probability maps are then rotated back and averaged.

\noindent\textbf{CNN for activation time estimation.} We develop a CNN-based neural network to predict the TOS curves for late activation time detection. Our network directly learns the relationship between time-series circumferential strain matrices of left ventricles and TOS curves via regression. In all experiments, we use standard convolutional layers with typical activation function ReLU and mean square error (MSE) as a loss. Since the activation for each segment happens no earlier than the time of an initial scan $t_0$, we employ a shifted version of the leaky ReLU~\cite{maas2013rectifier} to enforce such constraints in the last layer of network architecture:
\begin{align}
\sigma(x) = 
    \left\{\begin{matrix}
    x,                  &\text{ if \;} x \geq t_0\\ 
    -\alpha x + (1+\alpha)t_0,    &\text{otherwise}  
    \end{matrix}\right.
    \label{eq:shifted_leaky_relu}
\end{align}
Here, the input variable $x$ is an output of the fully connected layer in our network, and $\alpha$ is a positive constant (i.e., $\alpha=0.01$ is used in our experiments).


Similar to the segmentation network, we augment the training data for a better network performance. However, many commonly used augmentation methods (e.g. rotation, cropping, and flipping) are inapplicable in this task since they may introduce anatomically meaningless images. Therefore, we carefully apply cyclic translations along the segment axis.

\noindent\textbf{3D Visualization of activation Map.} After predicting the TOS curve from our network on all 2D slices from the same patient, we reconstruct a 3D activation map by simply interpolating TOS values on 3D surfaces of left ventricles reconstructed from the segmentation network. 

\section{Experiments}

The network training was performed on an Nvidia 2080Ti GPU with $11$ GB RAM over $1000$ epochs using an Adam optimizer~\cite{kingma2014adam}. We set the learning rate as $1E-4$ and the mini batch size as $300$.

\noindent\textbf{Data.} Data were acquired using a 1.5T MR scanner (Avanto, Siemens, Erlangen, Germany) with a four-channel phased-array radiofrequency coil. Cine DENSE was performed in 4 short-axis planes at basal, two mid-ventricular, and apical levels. Cine DENSE parameters included a temporal resolution of 17 ms, pixel size of $2.65 \times 2.65 \text{ mm}^2$ and slice thickness = $8$ mm. Displacement was encoded in two orthogonal directions and a spiral k-space trajectory was used with 6 interleaves per image. Other parameters included: field of view $= 240 \times 240 \text{ mm}^2$, displacement encoding frequency $k_e = 0.1 \text{ cycles/mm}$, flip angle $15^{\circ}$ and echo time $= 1.08 \text{ ms}$.

\noindent\textbf{Experiments.} We train the segmentation network on $12,415$ 2D images from $64$ patients, with $20\%$ of which are used for model validation. We then test on $5,255$ images from $44$ mixed subjects of patients and healthy volunteers. Four symmetric encoding and decoding blocks are used in the contracting and expanding path of the network, respectively. Each encoding block of the contracting path includes two consecutive sets of dilated convolutional layers with filter size $3\times 3$ and dilation rate as $2$, a batch normalization layer, and a rectified linear activation layer. Between each encoding block, pooling layers with step size of $3\times 3$ and stride as $2$ are applied to reduce the spatial dimension in all directions. Each decoding block contains two consecutive sets of deconvolutional layers with filter size $3\times 3$, a batch normalization layer, and a rectified linear activation layer.

To evaluate our network performance on the TOS curve prediction, we run experiments on $96$ cine DENSE MRIs collected from $15$ patients in total. A number of $1512$ MRIs after data augmentation from $12$ patients are used for training, and the rest for testing. Due to the fact that different data may contain different number of time frames, we zero-pad missing values to unify the dimension of strain matrix over time. We compare the predicted activation time of our method with both ground truth (manually delineated by experts) and the state-of-the-art algorithm~\cite{auger2017imaging}. 

\begin{figure}[ht!]
\centering
\begin{subfigure}{.46\linewidth}
  \centering
  \includegraphics[width=\linewidth]{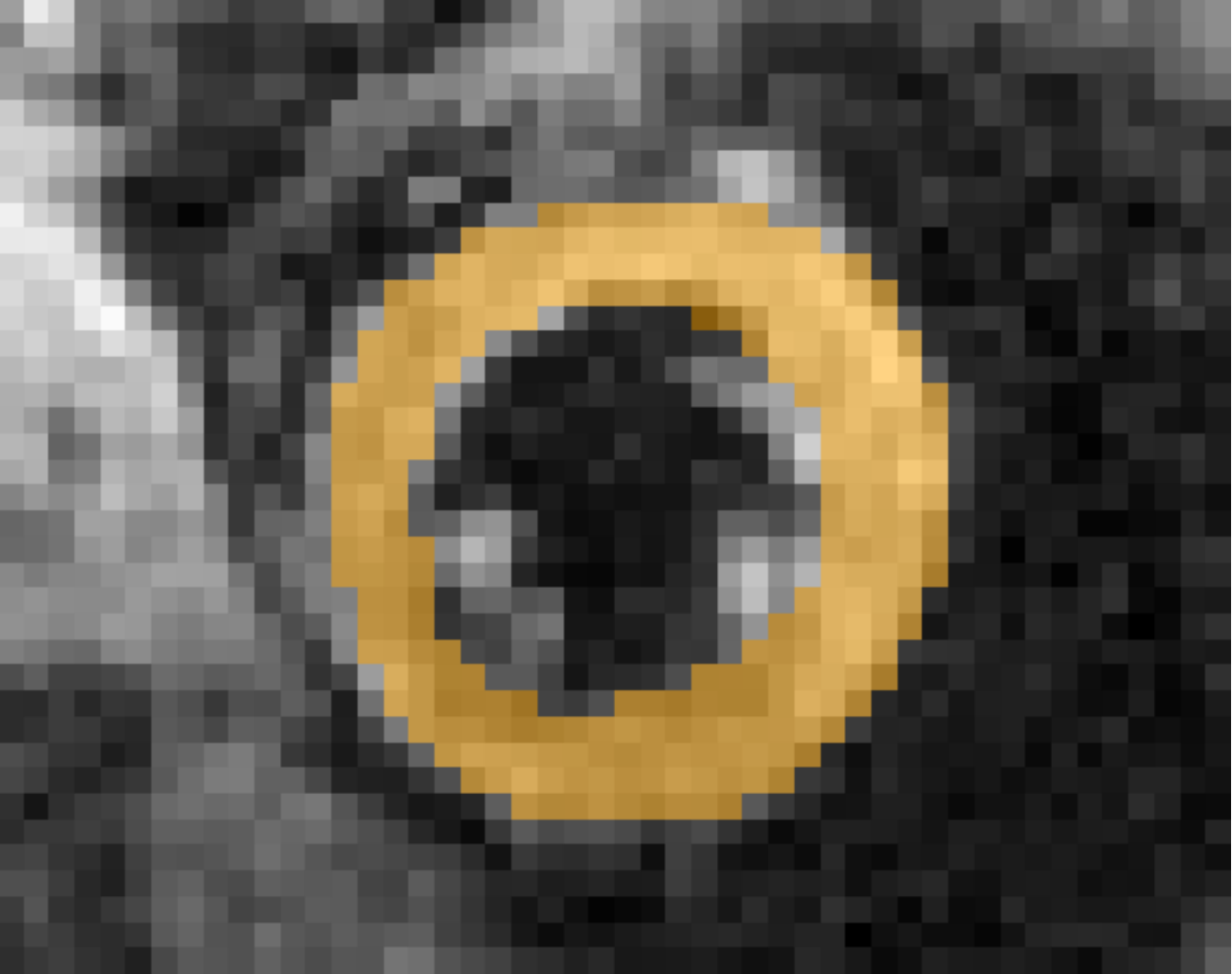}
  \label{fig:Seg-GT}
\end{subfigure}
\begin{subfigure}{.46\linewidth}
  \centering
  \includegraphics[width=\linewidth]{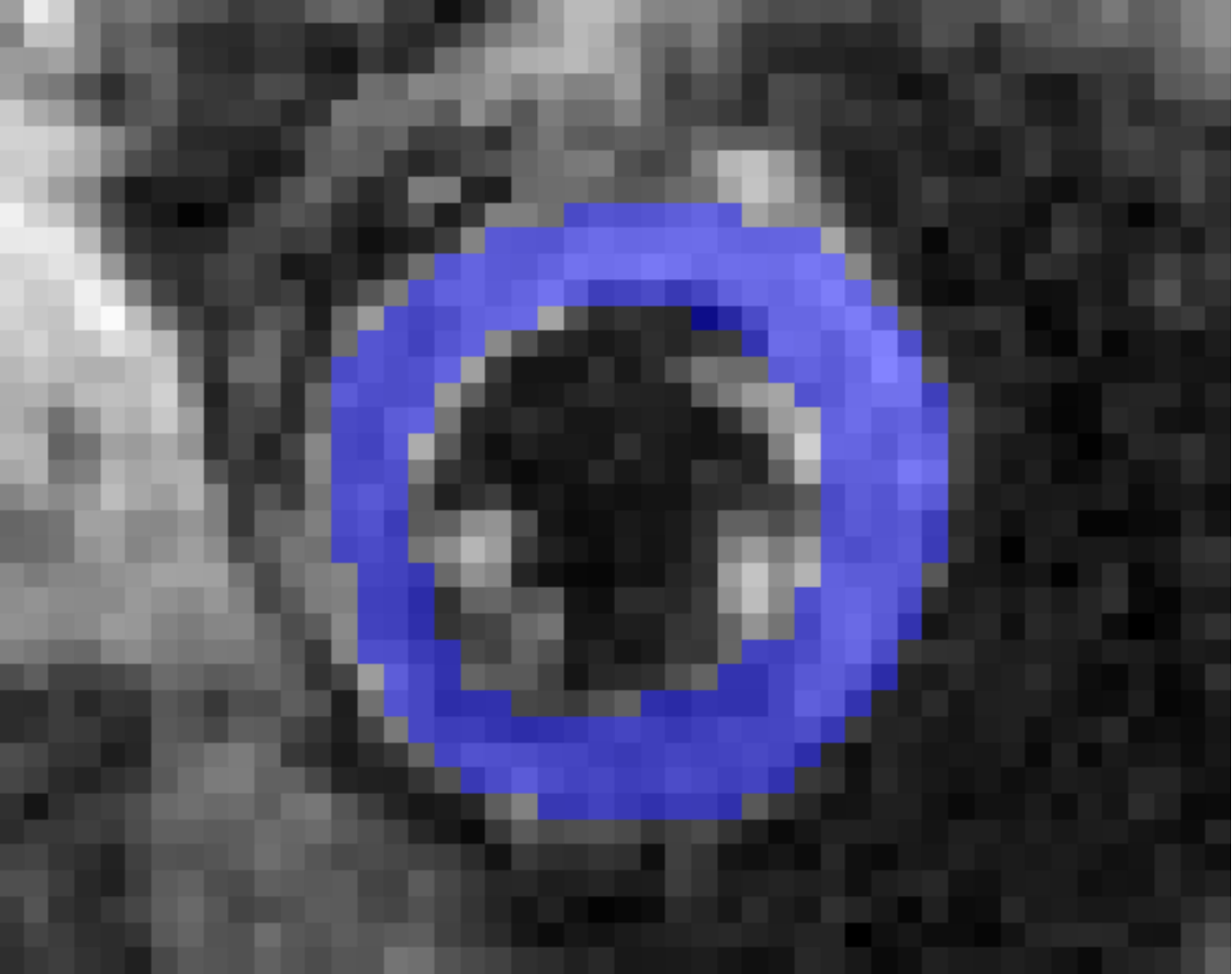}
  \label{fig:Seg-Esti}
\end{subfigure}
\caption{Left: estimated segmentation of left ventricle by our approach; Right: manual delineations.}
\label{fig:Seg}
\end{figure}

\noindent\textbf{Results.} Compared to manually labeled segmentation map, our network results in dice coefficient of left ventricles as $0.87 \pm 0.04$, a Hausdorff distance of $2.7 \pm 1$ pixel (equivalent to $5.94 \pm 2.2mm$), and a mean surface distance of $0.41 \pm 0.29$ pixels ($0.9 \pm 0.6 mm$). The computation time for determining the epicardial and endocardial contours for a single DENSE image is $0.16 \pm 0.02s$, $0.15 \pm 0.01s$, respectively. Fig.~\ref{fig:Seg} shows an example of estimated segmentation of left ventricle by our network vs. manually labeled ground truth. 

\begin{figure}[ht!]
\centering
\begin{subfigure}{.92\linewidth}
    \centering
    \begin{subfigure}{.49\linewidth}
      \centering
      \includegraphics[width=\linewidth]{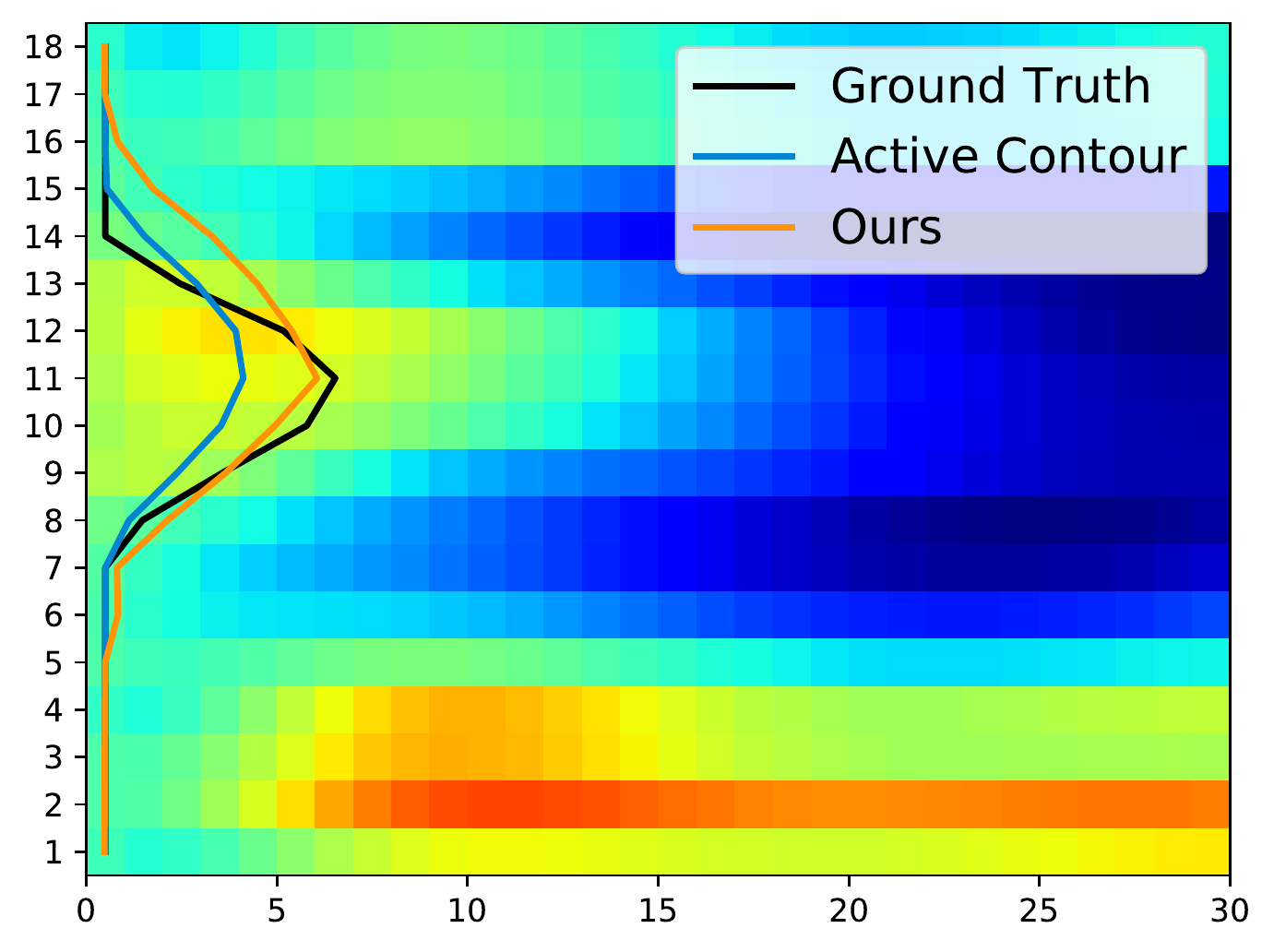}
      \label{fig:TOS_Pred_Test_2}
    \end{subfigure}%
    \begin{subfigure}{.49\linewidth}
      \centering
      \includegraphics[width=\linewidth]{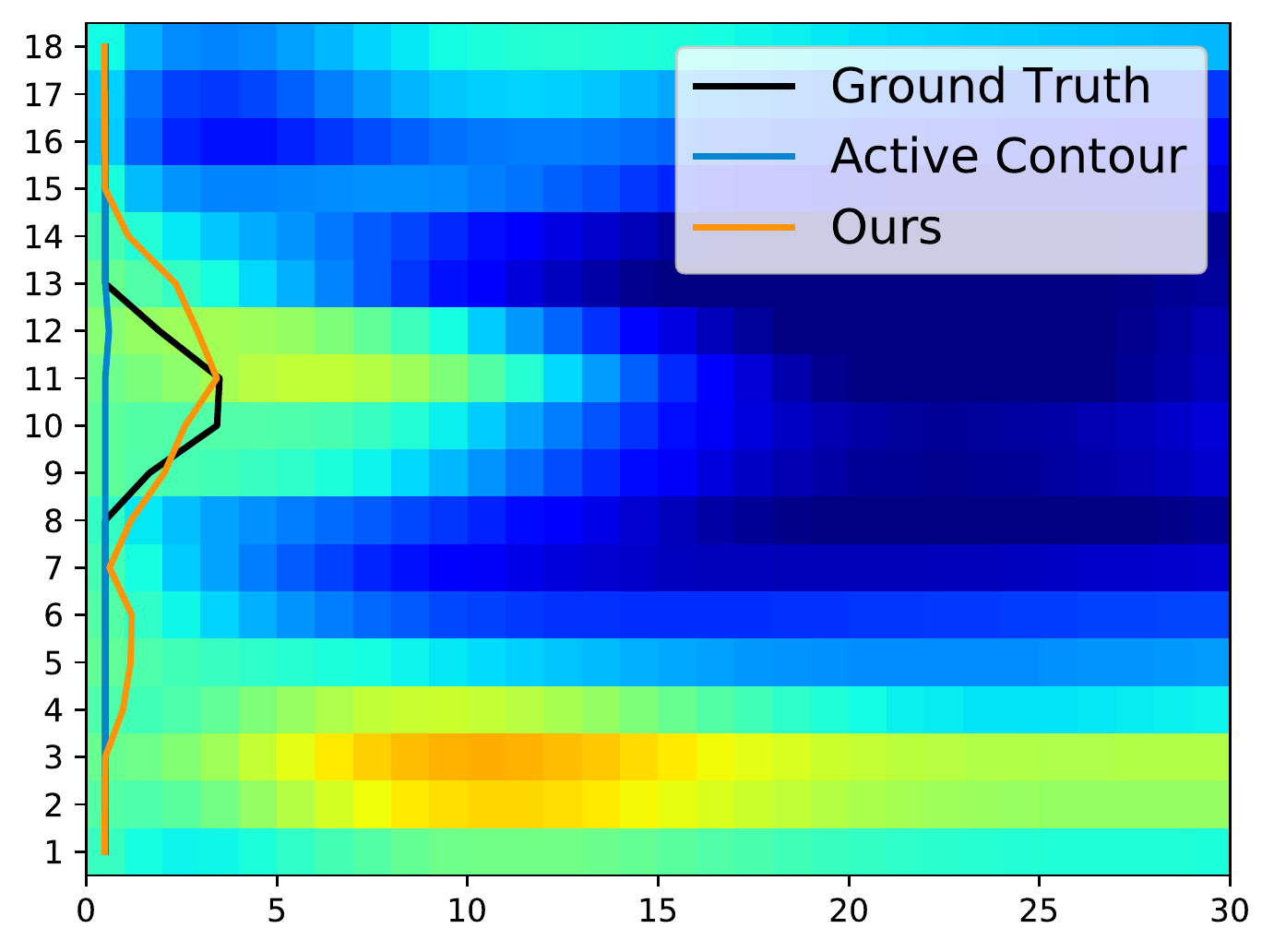}
      \label{fig:TOS_Pred_Test_3}
    \end{subfigure} \\
    \begin{subfigure}{.49\linewidth}
      \centering
      \includegraphics[width=\linewidth]{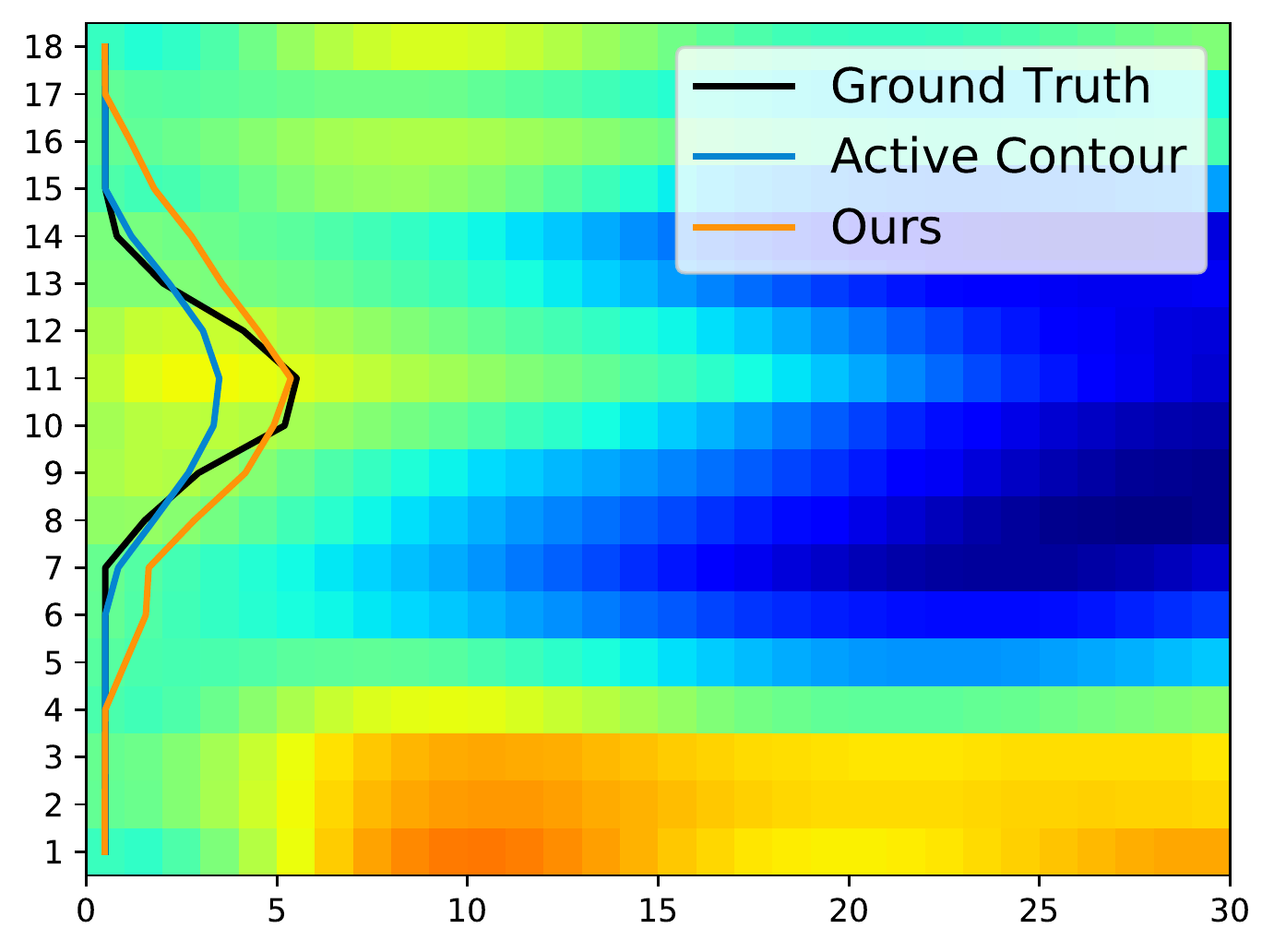}
      \label{fig:TOS_Pred_Test_5}
    \end{subfigure}
    \begin{subfigure}{.49\linewidth}
      \centering
      \includegraphics[width=\linewidth]{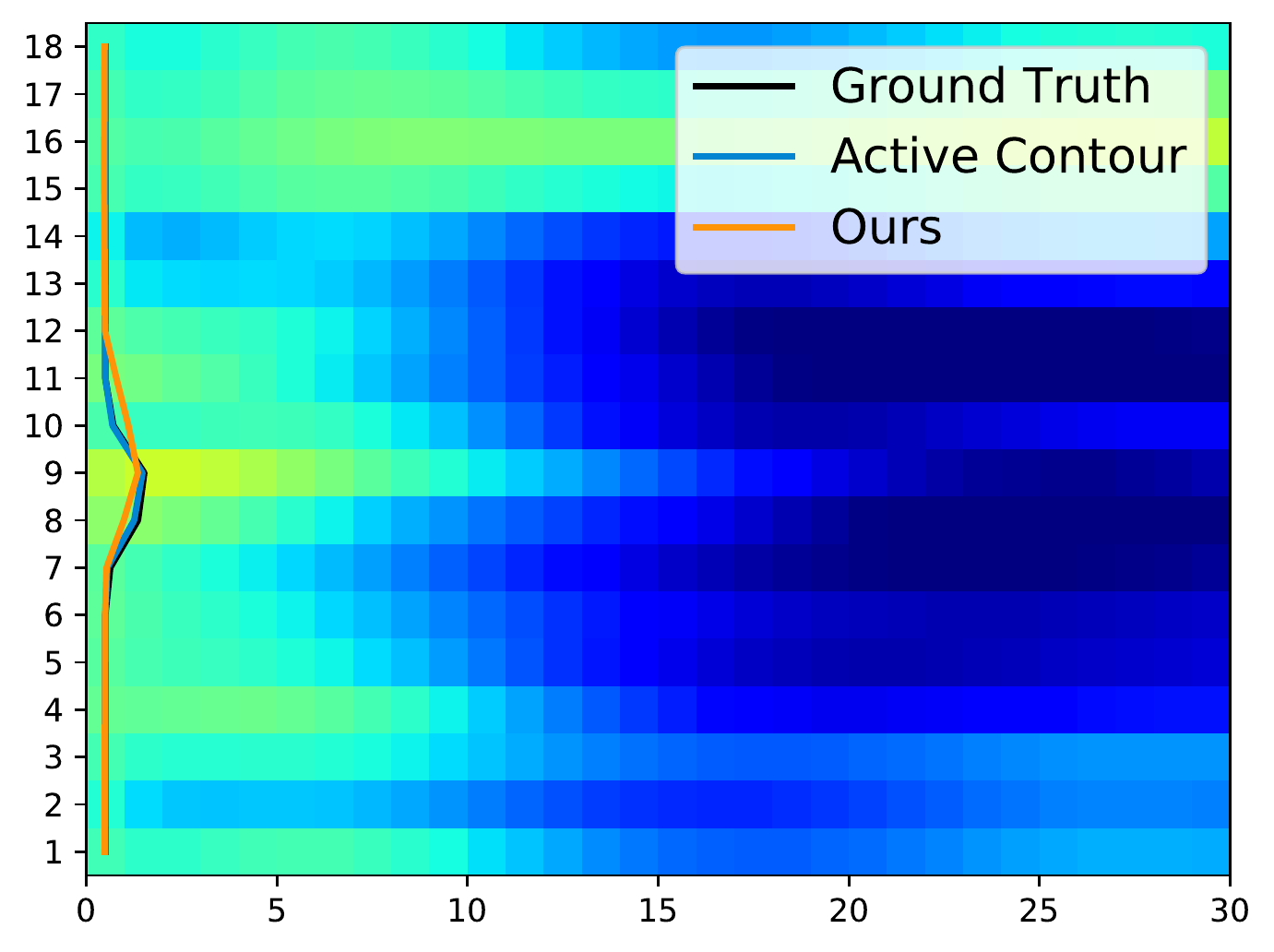}
      \label{fig:TOS_Pred_Test_6}
    \end{subfigure}
\end{subfigure}
\begin{subfigure}{.06\linewidth}
  \centering
  \includegraphics[width=\linewidth]{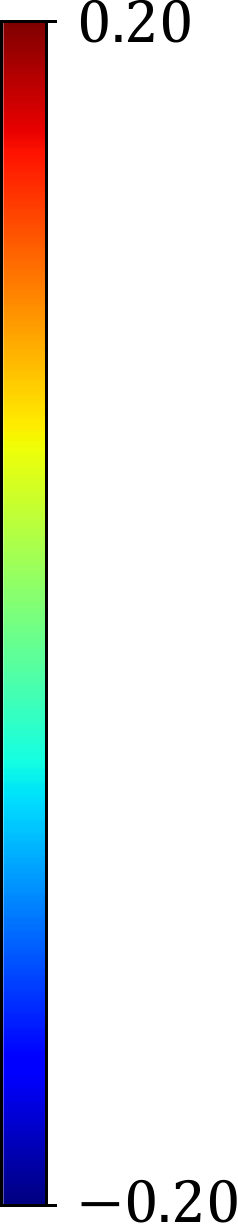}
\end{subfigure}
\caption{A comparison of TOS curves among ground-truth (black), baseline algorithm (blue), and our estimation (orange).}
\label{fig:TOS_Pred}
\end{figure}

Fig.~\ref{fig:TOS_Pred} displays examples of predicted TOS curves of test data. It shows that our approach outperforms the traditional algorithms based on active contours~\cite{auger2017imaging}. Aside from eliminating the need of parameter-tuning, our method runs an order of magnitude faster than the baseline algorithm when estimating the TOS ($0.001s$ vs. $0.676s$ per image).

\begin{figure}[ht!]
\centering
\begin{subfigure}{.46\linewidth}
  \centering
  \includegraphics[width=\linewidth]{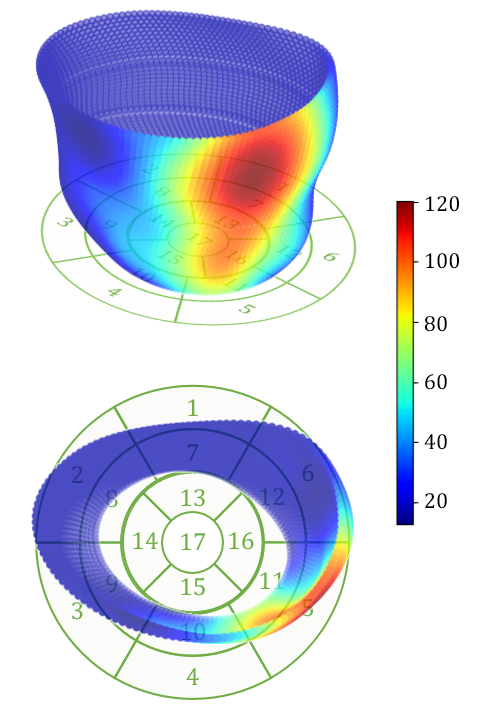}
  \caption{}
  \label{fig:3DMap-01-Pred-Comb}
\end{subfigure}
\begin{subfigure}{.46\linewidth}
  \centering
  \includegraphics[width=\linewidth]{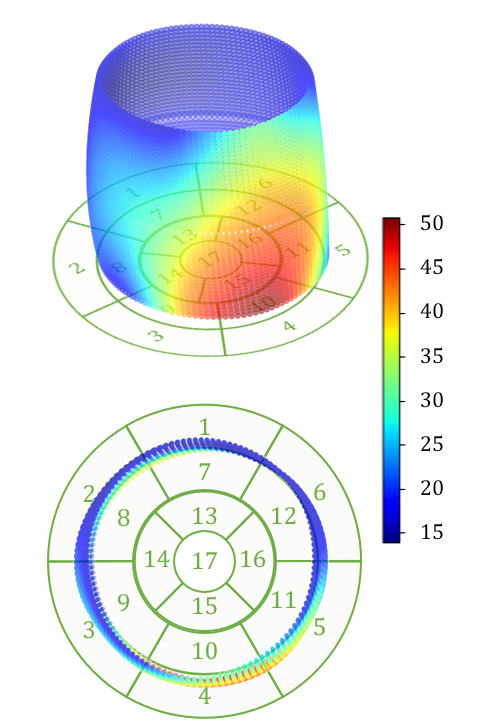}
  \caption{}
  \label{fig:3DMap-02-Pred-Comb}
\end{subfigure}
\begin{subfigure}{\linewidth}
  \centering
  \includegraphics[width=.8\linewidth]{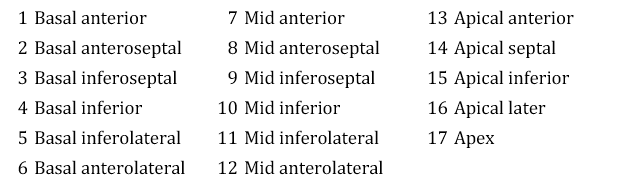}
  \label{fig:3DMap-Eye-Plot_Names}
\end{subfigure}
\caption{Examples of 3D activation Maps in column (a) and (b). Top panel: 3D view of activation maps; Bottom panel: bulls-eye plots.}
\label{fig:3DMap}
\end{figure}

Examples of reconstructed 3D activation maps from two patients are shown on the top panel of Fig.~\ref{fig:3DMap}. For visualization purpose, we employ the $17$-segment American Heart Association cardiac model, which has been shown to provide the best agreement with the available anatomical data~\cite{cerqueira2002american}. The Basal and mid-plane slices from the short-axis view are divided into six segments, apical into four segments, and apex into one segment (see the Bulls-eye plots on the bottom panel of Fig.~\ref{fig:3DMap}).  Fig. ~\ref{fig:3DMap-01-Pred-Comb} indicates that the most severe late activation (colored in dark red) locates at basel inferolateral section but mid inferior section in Fig.~\ref{fig:3DMap-02-Pred-Comb}. 

\section{Conclusion}
In this paper, we present a cascade network that automatically predicts delayed activation time of left ventricles from cardiac images. We develop a workflow that effectively integrates automatic segmentation of left ventricles, TOS prediction based on time-series circumferential strain analysis, and 3D visualization of cardiac activation maps. To the best of our knowledge, this is the first time that an end-to-end network is developed for fully automatic activation time detection of heart from cine DENSE MRIs. The ability of our work predicting better TOS to locate regions of severely delayed activation has great potential to improve the CRT response rate. While our experimental results are demonstrated on cine DENSE MRIs, our model is applicable to many other image modalities, such as ultrasound or spatial modulation of magnetization (SPAMM).

\bibliographystyle{abbrv}
\bibliography{main}

\end{document}